\documentclass[a4paper, twocolumn, 11pt, unpublished ]{quantumarticle}
\pdfoutput=1
\usepackage[utf8]{inputenc}
\usepackage[english]{babel}
\usepackage[T1]{fontenc}
\usepackage{amsmath}
\usepackage{amssymb}
\usepackage{amsthm}
\usepackage{tikz}
\usepackage{enumerate}
\usepackage{physics}

\usepackage{bbm} 

\usepackage{subcaption} 

\usepackage[numbers]{natbib} 

\usepackage{hyperref}
\hypersetup{
   breaklinks=true,   
   colorlinks=true,   
   pdfusetitle=true,  
}

\newcommand{\denop}{\mathcal{D}}
\newcommand{\hs}{\mathcal{H}}
\newcommand{\linop}{\mathcal{L}}
\DeclareMathOperator{\mytr}{Tr}

\newcommand{\E}{\mathcal{E}}

\newcommand{\id}{\mathbbm{1}}

\newcommand{\U}{\mathcal{U}}

\newcommand{\Enet}{\mathcal{E}_\mathrm{net}}
\newcommand{\Etar}{\mathcal{E}_\mathrm{tar}}
\newcommand{\EWalpha}{\mathcal{E}_{\mathrm{W},\alpha}}

\newcommand{\rhoin}{\rho_\mathrm{in}}
\newcommand{\rhotar}{\rho_\mathrm{tar}}
\newcommand{\rhoout}{\rho_\mathrm{out}}

\begin{document}

\title{The Impact of Architecture and Cost Function on Dissipative Quantum Neural Networks}

\author{Tobias C. Sutter}
\orcid{0009-0008-3179-7979}
\email{tobias.christoph.sutter@univie.ac.at}
\author{Christopher Popp}
\email{christopher.popp@univie.ac.at}
\author{Beatrix C. Hiesmayr}
\email{beatrix.hiesmayr@univie.ac.at}
\affiliation{University of Vienna, Faculty of Physics, Währingerstrasse 17, 1090 Vienna}

\maketitle

\begin{abstract}

Combining machine learning and quantum computation is a potential path towards powerful applications on quantum devices.
Regarding this, quantum neural networks are a prominent approach.
In this work, we present a novel architecture for dissipative quantum neural networks (DQNNs) in which each building block can implement any quantum channel, thus introducing a clear notion of universality suitable for the quantum framework.
To this end, we reformulate DQNNs using isometries instead of conventionally used unitaries, thereby reducing the number of parameters in these models.
We furthermore derive a versatile one-to-one parametrization of isometries, allowing for an efficient implementation of the proposed structure.
Focusing on the impact of different cost functions on the optimization process, we numerically investigate the trainability of extended DQNNs.
This unveils significant training differences among the cost functions considered.
Our findings facilitate both the theoretical understanding and the experimental implementability of quantum neural networks.

\end{abstract}

\section{Introduction} \label{sec:intro}


%
Classical machine learning (CML) and quantum computing are established computational paradigms.
While the former has already proven valuable in widely used applications like large language models, the theoretically promised advantages of the latter \cite{preskill_quantum_2012} are yet to be confirmed experimentally.
This is due to the experimental challenges accompanying the realization of quantum computers \cite{preskill_quantum_2018}.
Nonetheless, machine learning on quantum hardware, i.e., quantum machine learning (QML), promises several benefits, like reducing the complexity of specific machine learning algorithms \cite{cerezo_challenges_2022}.
One explicit model, often called dissipative quantum neural network (DQNN), has been proposed in Ref.~\cite{beer_training_2020}.
It can be understood as a straightforward quantization of classical feedforward artificial neural networks.
However, it manifestly does not contain nonlinearities, which are crucial for the universality (i.e., the ability to approximate any continuous function on a compact domain arbitrarily well) of its classical counterpart \cite{hornik_approximation_1991}.
In this regard, it is essential to emphasize that the notions of universality for CML and QML may differ, and obtaining a quantum advantage (e.g., speed-up over any classical algorithm) may be only one of many reasonable goals of QML \cite{schuld_is_2022}.
Furthermore, (linear) DQNNs are intriguing from a quantum information theoretic viewpoint as fundamental concepts like the Heisenberg uncertainty relation appear in their optimization process \cite{hiesmayr_quantum_2024}.
Despite the potential of QML, several factors affect the expressivity and trainability of these models:
Besides quantum hardware \cite{preskill_quantum_2018} and data-related factors \cite{schuld_effect_2021}, the cost function and network architecture \cite{cerezo_cost_2021, sharma_trainability_2022}, and entanglement within the network \cite{ortiz_marrero_entanglement-induced_2021, patti_entanglement_2021} crucially influence a model's performance.\\
%
%
\indent
The aim of this contribution is twofold: We first extend the conventional DQNN architecture so that each building block satisfies a specific notion of universality and subsequently focus on the impact of the cost function on the training process.
To avoid problems arising from the exponentially growing Hilbert space dimension, we concentrate on shallow DQNNs with a small output Hilbert space.
Thus, we provide small-scale results for the proposed extended architecture.
As an interesting use case, we mention that DQNNs with a single output qubit already allow us to infer three properties of the input state (one for each degree of freedom of the output state).
This is sufficient for specific quantum information processing tasks like determining the input state's purity or the concurrence \cite{wootters_entanglement_1998}.

%
%
The work is structured as follows.
We formally introduce DQNNs in Sec.~\ref{sec:conventional_DQNN} before reformulating them using isometries instead of unitaries in Sec.~\ref{sec:isometry_formulation_of_DQNNs}.
We derive a composite parametrization of isometries to leverage the resulting reduction of the variational parameters.
Sec.~\ref{sec:optimization_of_DQNNs} discusses two distinct training approaches: One based on random state sampling and the other on the Choi state of a quantum channel.
In Sec.~\ref{sec:extended_DQNNs}, we propose an extension of the conventional DQNN architecture based on considerations about the universality of DQNNs.
This ensures that each quantum perceptron has the power to implement a general quantum channel.
Sec.~\ref{sec:cost_funct} introduces various cost functions for mixed output and target states before we report our numerical results in Sec.~\ref{sec:numerical_results}.
Here, we conduct numerical simulations to assess the impact of different cost functions on the optimization process of a minimal extended DQNN.
Sec.~\ref{sec:numerical_results_random} is concerned with learning randomly sampled quantum channels, while Sec.~\ref{sec:numerical_results_werner} investigates the trainability of the Werner channel.
We conclude by discussing our results in Sec.~\ref{sec:conclusion}.

\section{Dissipative Quantum Neural Networks} \label{sec:DQNNs}

Dissipative quantum neural networks (DQNNs) are a straightforward quantization of classical feedforward artificial neural networks, where the artificial neurons are replaced by quantum systems.
Usually, these models' trainable weight and bias matrices are represented by variational unitary gates that are applied to the layers consecutively.
During the training phase, the unitary parameters are adjusted to optimize a given cost function that compares the network's output to a desired target output (supervised learning).
Due to the freedom of initializing the hidden and output layer neurons in fiducial quantum states, these unitaries can be considered as isometries.
This reduces the degrees of freedom and, thus, the computation effort required for the optimization procedure of DQNNs.
We derive a versatile one-to-one parametrization of isometries from the composite parametrization of the unitary group $\mathcal{U}(d)$ \cite{spengler_composite_2010, spengler_composite_2012}.
The details can be found in App.~\ref{app:comp_param_for_isometries}.

Moreover, we define a DQNN as \textit{quantum channel universal} if it can implement any completely positive and trace-preserving (CPTP) map from the input to the output state.
This allows for a standardization of DQNNs and a meaningful performance comparison.
These considerations lead to an extended DQNN architecture where each building block naturally implements a general CPTP map.
In contrast to conventional DQNNs, a minimal version of our modified architecture (comprised of three neurons) is quantum channel universal.
The isometry viewpoint also gives a straightforward interpretation of the training process: The network aims to learn the Stinespring representation of a target quantum channel by adjusting its isometry degrees of freedom.

\subsection{Conventional DQNNs} \label{sec:conventional_DQNN}

\begin{figure*}[ht]
    \begin{center}
    \includegraphics[width=1\linewidth, keepaspectratio]{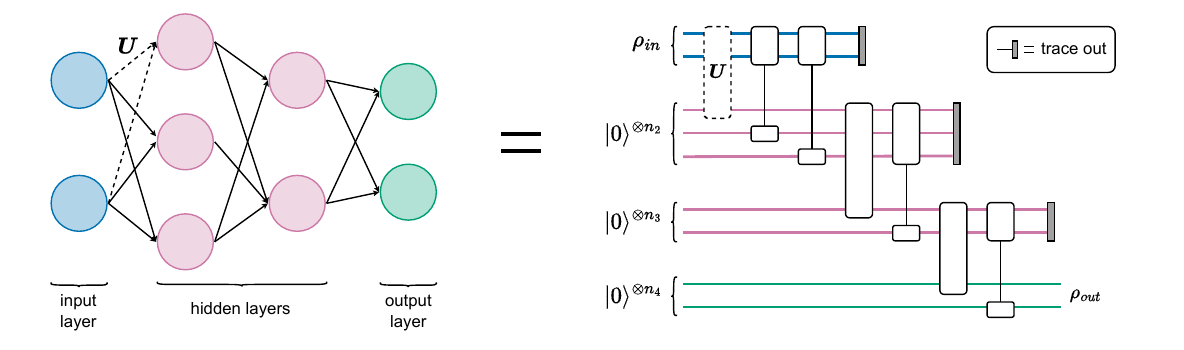}%
    \end{center}
    \vspace{-5mm}
    \caption[]{
    \label{fig:dqnn_without_ancilla_layers}
    Conventional dissipative quantum neural network (DQNN) with $N=9$, $n_1=n_3=n_4=2$, and $n_2=3$.
    It consists of input (blue), hidden (violet), and output layers (green).
    In the unitary formulation, the hidden and output layers are initialized in fiducial states.
    The black arrows in the left diagrams represent the unitary perceptrons and, thus, the information flow.
    Dashed lines indicate the first unitary $U = U_{1}^{(1,2)}$, while the full unitary is given by $U_{2}^{(3,4)} U_{1}^{(3,4)} U_{2}^{(2,3)} U_{1}^{(2,3)} U_{3}^{(1,2)} U_{2}^{(1,2)} U_{1}^{(1,2)}$.
    Hence, the parameter $k \in \{1,\dots,K_{\ell+1}\}$ that counts the neurons of each layer in $\eqref{eq:unitary_dqnn_general}$ increases from the top to the bottom neurons.
    After the unitaries are applied, the input and hidden layers are traced out, leaving the network in the output state \eqref{eq:output_state_dqnn}.
    In the quantum circuit diagram on the right, each horizontal line corresponds to one neuron (qudit), and time increases from the left to the right.
    The boxes represent the unitary perceptrons, successively applied in a specific order to adjacent layers.
    }
\end{figure*}

As introduced in Ref.~\cite{beer_training_2020}, the conventional architecture for DQNNs aims to mimic classical feedforward artificial neural networks:
The artificial neurons are represented by $d$-dimensional quantum systems called qudits, and unitary interactions represent the weight and bias matrices.
Choosing an architecture requires arranging these $N$ quantum neurons into $L$ layers, as visualized in Fig.~\ref{fig:dqnn_without_ancilla_layers}.
Layers 1 and $L$ constitute the input and output layers, respectively, and layers 2 to $L-1$ represent the hidden layers.
Each layer $\ell\in\{1,\dots,L\}$ consists of $n_\ell$ neurons.
Furthermore, we can formally assign a Hilbert space $\hs_\ell = \bigotimes_{i=1}^{n_\ell} \hs_{\ell}^{(i)}$ to each layer $\ell \in\{1, \dots,L\}$ of the network, where $\hs_{\ell}^{(i)}$ is the Hilbert space of the $i$th neuron in layer $\ell$.

\subsubsection{Unitary Formulation}
Initially, layers $\ell \geq 2$ are prepared in a fiducial state, e.g., the computational basis state $|0\rangle^{\otimes n_{\ell}} \in \hs_\ell$, and layer 1 holds a generally mixed input quantum state $\rhoin \in \denop(\hs_1)$.
The set $\denop(\hs_1)$ denotes the set of positive semi-definite linear operators mapping $\hs_1$ into itself and satisfying $\mytr(\rhoin)=1$.
Analogous to the weight and bias matrices in classical networks, neurons in adjacent layers ($\ell$ and $\ell+1$) of DQNNs are connected by variational unitary transformations $U_{k}^{(\ell,\,\ell+1)}$, called (quantum) perceptrons.
Here, $\ell \in \{1, \dots, L-1\}$ and $k \in \{1, \dots, K_{\ell+1}\}$, where $K_{\ell+1}$ is the total number of unitaries connecting layers $\ell$ and $\ell+1$.
For simplicity, we assume that each unitary $U_{k}^{(\ell,\,\ell+1)}$ acts on all neurons of layer $\ell$ (global perceptron) and one neuron of layer $\ell+1$.
Hence, $K_{\ell+1} = n_{\ell+1}$ and $U_{k}^{(\ell,\,\ell+1)} \in \mathcal{U}(\hs_{\ell+1}^{(k)} \otimes \hs_\ell )$, where $\mathcal{U}(\hs)$ denotes the set of unitary matrices acting on $\hs$.
After all unitaries are applied, layers $1$ to $L-1$ are traced out, yielding output state $\rhoout \in \denop(\hs_{L})$, given by
\begin{align} \label{eq:output_state_dqnn}
    \rhoout = \mytr_{1,\dots,L-1} \left[\, U\, (|\mathbf{0}\rangle\langle\mathbf{0}|_{L,\dots,2} \,\otimes \, \rhoin )\, U^\dagger \,\right] \;,
\end{align}
where
\begin{align}
    |\mathbf{0}\rangle_{L,\dots,2} &= |0\rangle_L \otimes |0\rangle_{L-1} \otimes \dots \otimes |0\rangle_2 \;, \\
    U &= \prod_{\ell=1}^{L-1} \left( \prod_{k=1}^{K_{\ell'+1}} U_{1 - k + K_{\ell'+1}}^{(\ell' ,\, \ell'+1)} \right) \;, \label{eq:unitary_dqnn_general}
\end{align}
where $\ell' = L-\ell$, and $|0\rangle_\ell = \bigotimes_{i=1}^{n_\ell} |0\rangle_{\ell}^{(i)}$ with $|0\rangle_{\ell}^{(i)} \in \hs_{\ell}^{(i)}$.
The first product in \eqref{eq:unitary_dqnn_general} concerns the layers, while the second regards the unitaries within one layer.
Two remarks are in order.
First, our choice for arranging the Hilbert space throughout this work is reversed in the sense that we consider the total Hilbert space of the DQNN as $\hs = \hs_L \otimes \hs_{L-1} \otimes \dots \otimes \hs_2 \otimes \hs_1$.
This is in preparation for using the composite parametrization of isometries for which this ordering is essential (cf. App.~\ref{app:comp_param_for_isometries}).
Second, we define the product in \eqref{eq:unitary_dqnn_general} as $\prod_{i=1}^{n} A_i = A_1 \cdot A_2 \cdots A_n$.
Thereby, we ensure that the DQNN applies the unitaries layer-wise and ordered according to the label $k$ (cf. Fig.~\ref{fig:dqnn_without_ancilla_layers}).
This is important because they generally do not commute within each layer.

We note in passing that this quantum machine learning ansatz crucially differs from its classical counterpart in that it does not involve any nonlinearities, which are essential for the universality property of classical feedforward artificial neural networks.
However, linear transformations are sufficient for the notion of universality we consider in Sec.~\ref{sec:extended_DQNNs}.

\subsubsection{From Unitaries to Isometries} \label{sec:isometry_formulation_of_DQNNs}

A different and computationally advantageous perspective on DQNNs can be adopted by considering the perceptrons not as unitary transformations but as isometries.
To do so, the neurons in the layers $\ell \in \{2, \dots, L\}$ are not initialized in a fiducial state.
Instead, the network's initial state is simply $\rhoin \in \denop(\hs_1)$.
Subsequently, the perceptron isometries $V_k^{(\ell, \ell+1)} := U_k^{(\ell,\ell+1)} |0\rangle^{(k)}_{\ell+1} \in \mathrm{Iso}(\hs_\ell, \, \hs_{\ell+1}^{(k)} \otimes \hs_\ell)$ get applied sequentially, thus each enlarging the network's Hilbert space $\hs$ by one neuron.
The output state is
\begin{align} \label{eq:output_state_dqnn_isometric_formulation}
    \rhoout = \mytr_{1,\dots,L-1} \left[\, V\, \rhoin \, V^\dagger \,\right] \;,
\end{align}
where
\begin{align} \label{eq:isometry_dqnn_general}
    V &= \prod_{\ell=1}^{L-1} \left( \prod_{k=1}^{K_{\ell'+1}} V_{1 - k + K_{\ell'+1}}^{(\ell',\, \ell' + 1)} \right) \;,
\end{align}
with $\ell' = L-\ell$.
One advantage of this formulation is that it illuminates the DQNN's implementation of a completely positive and trace-preserving (CPTP) map $\Enet : \denop(\hs_1)\rightarrow\denop(\hs_L)$, where $V$ can be viewed as the network's Stinespring isometry \cite{watrous_theory_2018}.
We can therefore write $\rho_\mathrm{out} = \mathcal{E}_\mathrm{net}(\rho_\mathrm{in})$, describing every transformation that is theoretically possible for DQNNs.
Another advantage is that isometries have fewer degrees of freedom than unitaries.
Hence, this reformulation also reduces the number of parameters to optimize during the network's training phase (described below).
In particular, a unitary perceptron acting on $d$-dimensional input and hidden layer qudits has $d^4$ free parameters, while an isometry perceptron acting on the same qudits only has $d^2(2d-1)$.
Already for $d=4$, this more than halves the number of parameters to optimize.

However, a suitable variational parametrization of isometries is required to exploit this.
Based on the composite parametrization (CP) of the unitary group \cite{spengler_composite_2010, spengler_composite_2012}, we derive a corresponding one-to-one parametrization of isometries in App.~\ref{app:comp_param_for_isometries}.
As a result, any isometry $V_\mathrm{CP} \in \mathrm{Iso}(\hs_1,\hs_2)$ can be written as
\begin{align}
    V_\mathrm{CP} = \left[\prod_{m=0}^{d_1-1} \prod_{n=m+1}^{d_{2}-1} \Lambda_{m,n} \right]\left[\prod_{l=0}^{d_1-1} e^{i P_l \lambda_{ll}} \right] \, \id_{d_{2}\times d_1} \;, \label{eq:comp_par_iso_main_text}
\end{align}
where $d_\ell=\dim(\hs_\ell)$, $\{|i\rangle_\ell\}_{i=0}^{d_\ell-1}$ is a basis of $\hs_\ell$, and
\begin{align}
    \id_{d_{2}\times d_1} &= \sum_{i=0}^{d_1-1} |i\rangle_{2} \langle i|_1 \;,\\
    P_n &= |n\rangle_{2} \langle n|_{2} \;,\\
    Y_{m,n} &= -i |m\rangle_{2}\langle n|_{2} + i |n\rangle_{2}\langle m|_{2} \; ,\\
    \Lambda_{m,n} &= e^{i\, P_n \lambda_{n,m}} e^{i\, Y_{m,n} \lambda_{m,n}} \;.
\end{align}
The set $\{\lambda_{m,n} \,|\, 0\leq m,n< d_2, \, m< d_1 \,\vee\, n< d_1\}$ contains the $2d_1 d_2 - d_1^2$ parameters of $V_\mathrm{CP}$.

\subsubsection{Gradient Optimization} \label{sec:optimization_of_DQNNs}
The standard procedure for training the network toward implementing a desired target transformation $\mathcal{E}_{\mathrm{tar}} : \denop(\hs_1)\rightarrow\denop(\hs_L)$ involves sampling a set of input states $\{\rho_{\mathrm{in}}^{(i)}\}_{i=1}^{N_t}$.
This random element can speed up the optimization process, similar to stochastic gradient descent in classical machine learning.
However, the geometry of quantum state space is non-unique \cite{bengtsson_geometry_2006}, so this scheme can suffer from choosing the ``wrong'' sampling method.
For each element of the input state set, the corresponding network output state $\rhoout^{(i)} = \Enet (\rhoin^{(i)})$ and target output state $\rhotar^{(i)} = \Etar (\rhoin^{(i)})$ are computed.

It is imperative that $\mathcal{E}_{\mathrm{tar}}$ is (close to) a CPTP map.
Otherwise, the DQNN will inevitably fail in the training process as $\Enet$ is necessarily a quantum channel, and the perfect network satisfies $\Enet(\sigma)=\Etar(\sigma)$ for all $\sigma \in \denop(\hs_1)$.
Hence, a good strategy to avoid trainability issues is to ensure that the target transformation $\Etar$ represents a quantum channel, i.e., is linear and CPTP.

To evaluate how well the network reproduces $\mathcal{E}_{\mathrm{tar}}$, a cost/loss function $C:\denop(\hs_L)\times\denop(\hs_L) \rightarrow \mathbbm{R}$ is applied to each element of $\{ ( \rho_\mathrm{out}^{(i)} , \rho_\mathrm{tar}^{(i)} )\}_{i=1}^{N_t}$.
The total cost of the network is the average cost over all training states,
\begin{align} \label{eq:cost_total}
    C_\mathrm{tot} = \frac{1}{N_t} \sum_{i=1}^{N_t} C(\rho_\mathrm{tar}^{(i)}, \, \rho_\mathrm{out}^{(i)}) \;.
\end{align}
The function $C$ is usually a similarity or distinguishability measure on the output state space.
We discuss potential candidates in Sec.~\ref{sec:cost_funct} and their impact on the training in Sec.~\ref{sec:numerical_results}.

Once a cost function is chosen, the network trains by updating the variational isometry parameters $\{\lambda_\mu\}_\mu$ according to gradient descent (if $C$ is a distinguishability measure) or gradient ascent (if $C$ is a similarity measure).
This requires taking the derivative of \eqref{eq:cost_total} with respect to every $\lambda_\mu$,
\begin{align} \label{eq:gradient_general_cost}
    \frac{\partial C_\mathrm{tot}}{\partial \lambda_\mu} = \frac{1}{N_t} \sum_{i=1}^{N_t} \frac{\partial}{\partial \lambda_\mu} C(\rho_\mathrm{tar}^{(i)}, \, \rho_\mathrm{out}^{(i)}) \;,
\end{align}
and adjust the network's parameters according to, e.g., the ADAM optimizer \cite{kingma_adam_2017}.
Repeating this feedback loop of computing the network's output state for each training input state and updating the isometry parameters leads to a (local) optimum in the cost function landscape.
We call this scheme random state training.

Due to the fact that DQNNs can only realize quantum channels, a different optimization method can be considered.
It does not rely on random state sampling but the Choi representation of quantum channels \cite{choi_completely_1975, jamiolkowski_linear_1972}.
The maps $\Enet$ and $\Etar$ are identified with their respective Choi state \cite{watrous_theory_2018}
\begin{align}
    J(\mathcal{E}_\mathrm{net/tar}) = \frac{1}{d_1} \sum_{i,j=0}^{d_1-1} \mathcal{E}_\mathrm{net/tar} \left(|i\rangle_1\langle j|_1\right) \otimes |i\rangle_1\langle j|_1 \;.
\end{align}
Operationally, the state $J(\Enet)$ can be created by sending one half of the maximally entangled state $|\Omega\rangle = 1/\sqrt{d_1} \sum_{i=0}^{d_1-1} |i\rangle \otimes |i\rangle$ through the network.
Because the Choi representation is unique, the DQNN perfectly represents the target transformation if and only if $J(\mathcal{E}_\mathrm{net}) = J(\mathcal{E}_\mathrm{tar})$.
Thus, defining a cost function $C:\denop(\hs_L\otimes\hs_1)\times\denop(\hs_L\otimes\hs_1) \rightarrow \mathbbm{R}$, we can optimize the DQNN by computing
\begin{align}
    \frac{\partial C(J(\mathcal{E}_\mathrm{tar}),\, J(\mathcal{E}_\mathrm{net}))}{\partial \lambda_\mu} \;,
\end{align}
and using gradient optimization as before.
We refer to this method as Choi training.
The drawbacks are that the target channel must be entirely known, and the cost function acts on a larger Hilbert space.
However, it does not suffer from a potentially unsuitable sampling of (finitely many) input quantum states and thus allows more objective trainability statements.
Therefore, we use it to benchmark the performance of different cost functions in Sec.~\ref{sec:numerical_results}.

\subsection{Extended DQNNs} \label{sec:extended_DQNNs}

\begin{figure*}[ht]
    \begin{center}
    \includegraphics[width=0.9\linewidth, keepaspectratio]{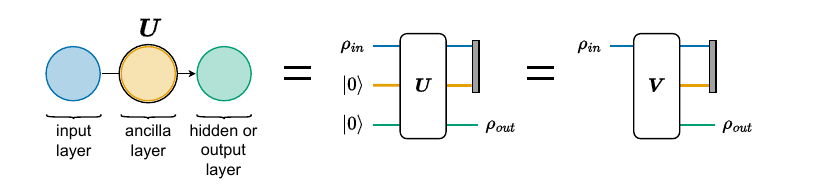}%
    \end{center}
    \vspace{-5mm}
    \caption[]{
    \label{fig:minimal_universal_dqnn}
    The minimal version of an extended DQNN consists of three neurons and can be viewed as the prototype for perceptrons of larger extended DQNNs.
    The black ring and arrow on the left denote the perceptron.
    It can learn any quantum channel from the input to the hidden/output neurons if the dimensions of the ancilla and the hidden/output neurons coincide.
    The middle and right figures show the quantum circuit of the perceptron in the unitary and isometry formulation, respectively, for which we have $V = U (|0\rangle_3 \otimes |0\rangle_2 \otimes \id_1)$.
    For general CPTP maps, $V$ does not factorize, i.e., $V \neq V_1^{(2,3)} \, V_1^{(1,2)}$.
    }
\end{figure*} 

It is clear from \eqref{eq:output_state_dqnn_isometric_formulation} that DQNNs implement a CPTP map from $\denop(\hs_1)$ to $\denop(\hs_L)$.
Consequently, the most general learnable transformation $\mathcal{E}_\mathrm{tar}$ is also of this kind.
This leads us to regard a DQNN as \textit{quantum channel universal} if it can realize any CPTP map $\E:\denop(\hs_1)\rightarrow\denop(\hs_L)$.
If the DQNN can realize such a map only approximately, we say it is a universal quantum channel approximator.
Note that this definition only covers linear maps from the input to the output state; classical post-processing is needed to obtain nonlinear functions of the input state.
Furthermore, it contrasts universal quantum computation, which only regards approximating unitary transformations.

Conventional DQNNs (Sec.~\ref{sec:conventional_DQNN}) do not necessarily have a structure that enables quantum channel universality.
Take, e.g., a network consisting of one input, one hidden, and one output layer neuron.
There are two perceptrons in such a network, and according to \eqref{eq:output_state_dqnn_isometric_formulation} and \eqref{eq:isometry_dqnn_general}, the output state is
\begin{align}
    \rho_\mathrm{out} =  \mytr_{1,2} \left[\,V \, \rhoin \, V^\dagger \,\right] \;,
\end{align}
where $V=  V_1^{(2,3)} \, V_1^{(1,2)}$.
However, a Stinespring isometry $V_\mathcal{E}$ of a quantum channel $\mathcal{E}:\denop(\hs_1)\rightarrow\denop(\hs_3)$ can generally not be written as the product of two isometries, i.e., $V_\mathcal{E} \neq V_1^{(2,3)} \, V_1^{(1,2)}$.
Hence, this network is not quantum channel universal.

For this reason, we extend the input-hidden-output layer structure of DQNNs by adding ancilla layers.
Every perceptron adds to the network not only one hidden or output neuron but also an ancilla neuron, which is subsequently traced out (see Fig.~\ref{fig:minimal_universal_dqnn}).
Hence, the isometries are given by $V_k^{(\ell, \ell+1, \ell+2)} \in \mathrm{Iso}(\hs_\ell, \, \hs_{\ell+2}^{(k)} \otimes \hs_{\ell+1}^{(k)} \otimes \hs_\ell)$ and the output state $\rhoout$ results from \eqref{eq:output_state_dqnn_isometric_formulation} together with
\begin{align} \label{eq:isometry_dqnn_extended}
    V &= \prod_{\ell=1}^{(L-1)/2} \left( \prod_{k=1}^{K_{\ell'+2}} V_{1 - k + K_{\ell'+2}}^{(\ell',\, \ell'+1, \, \ell'+2)} \right) \;,
\end{align}
where $\ell' = L - 2\ell$.
The additional degree of freedom ensures that a minimal network consisting of an input, an ancilla, and an output layer connected by a single perceptron is quantum channel universal, provided that $\dim(\hs_2)=\dim(\hs_3)$.
The isometry viewpoint allows a straightforward interpretation of the training process: Given a target quantum channel, the network aims to learn its Stinespring representation.

\begin{figure*}[ht]
    \begin{center}
    \includegraphics[width=1\linewidth, keepaspectratio]{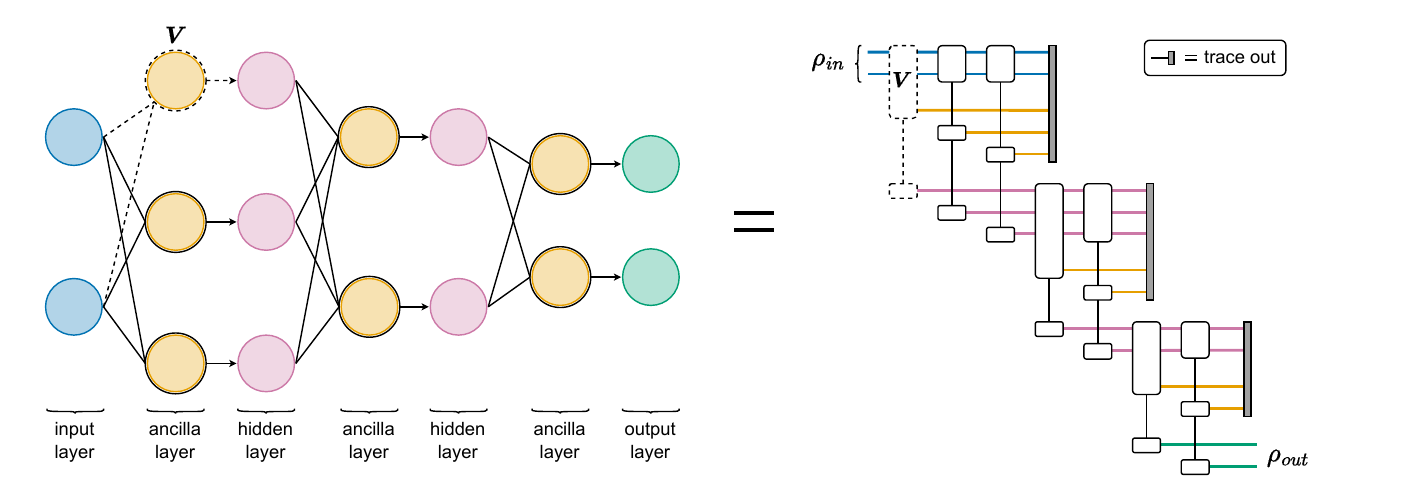}%
    \end{center}
    \vspace{-5mm}
    \caption[]{
    \label{fig:dqnn_with_ancilla_layers}
    An extended version of the network in Fig.~\ref{fig:dqnn_with_ancilla_layers}, consisting of 16 instead of 9 neurons and thus almost doubling the size.
    It comprises input (layer 1; blue), ancilla (layers 2, 4, and 6; gold), hidden (layers 3 and 5; violet), and output layers (layer 7; green).
    This comes with the benefit that every perceptron can implement a general CPTP map.
    The dashed lines indicate the first isometry perceptron $V=V_{1}^{(1,2,3)}$, while the full isometry \eqref{eq:isometry_dqnn_extended} is given by $V_{2}^{(5,6,7)} V_{1}^{(5,6,7)} V_{2}^{(3,4,5)} V_{1}^{(3,4,5)} V_{3}^{(1,2,3)} V_{2}^{(1,2,3)} V_{1}^{(1,2,3)}$.
    The quantum circuit diagram on the right utilizes the isometric formulation of extended DQNNs.
    }
\end{figure*}

This minimal extended DQNN can be considered the blueprint for the perceptrons of larger networks comprising multiple layers with more than one neuron each (see Fig.~\ref{fig:dqnn_with_ancilla_layers}).
Consequently, each perceptron of an extended DQNN is quantum channel universal.
Note, however, that this does not ensure that the whole network also has this property.

\section{Choices of Cost Functions} \label{sec:cost_funct}

The trainability of DQNNs depends on the cost function used in the optimization process \cite{cerezo_cost_2021, sharma_trainability_2022}.
Thus, a suitable cost function is essential for designing a useful DQNN.
In principle, one may choose any reasonable function $C:\denop(\hs_L)\times\denop(\hs_L) \rightarrow \mathbbm{R}$.
However, we focus on distance and similarity measures on $\denop(\hs_L)$ as they align with the usual ``cost'' or ``reward'' imposed for assessing the network's output.
In this section, we present several candidates for $C$ that are applicable to mixed output and target states of a DQNN.
Special attention is paid to experimental measurability and information-theoretic interpretation of the presented quantities.
Furthermore, we need the gradient of $C$ to optimize a DQNN using gradient descent/ascent.
This involves taking derivatives of $C$ with respect to the variational parameters of the network.
We present analytical expressions for this in App.~\ref{app:derivatives_cost_funct} whenever possible.

Typically, \textit{distance measures} $D:\denop(\hs)\times\denop(\hs)\rightarrow\mathbbm{R}$ on the space of density matrices are defined by the following properties:
$D$ must be nonnegative ($D(\rho,\sigma)\geq0$), symmetric ($D(\rho,\sigma)=D(\sigma,\rho)$), zero if and only if the states are equal ($D(\rho,\sigma)=0 \Leftrightarrow \rho=\sigma$), and satisfy the triangle inequality ($D(\rho,\sigma) \leq D(\rho,\chi)+D(\chi,\sigma)$).
Additionally, Ref.~\cite{vedral_quantifying_1997} proposes that a quantum distance measure should satisfy the so-called \textit{data-processing inequality}
\begin{align}
    D(\E(\rho),\E(\sigma)) \leq D(\rho,\sigma) \;,
\end{align}
where $\E$ is any CPTP map.
This allows using $D$ to quantify entanglement in a meaningful way.

Nevertheless, dropping some of these properties in favor of a clear operational interpretation can help solve specific problems.
In this case, one considers \textit{divergences}, which are not required to be symmetric or satisfy the triangle inequality.
The essential property is that they satisfy the data-processing inequality.
They find meaning, e.g., in asymmetric hypothesis testing scenarios, by quantifying how distinguishable one state is from another.

Similarly, \textit{fidelities} are a pivotal similarity measure between two quantum states.
By definition, every fidelity function $F$ must satisfy a set of axioms \cite{jozsa_fidelity_1994}.
One of them demands that if $F(\rho, \sigma)$ is a fidelity for $\rho, \sigma \in \denop(\hs_L)$, it reduces to $F(\rho, |\psi\rangle\langle\psi|) = \langle\psi|\rho|\psi\rangle$ if $\sigma=|\psi\rangle\langle\psi|$ is a pure state.
Hence, $F$ generalizes the notion of the transition probability of two pure states to the mixed case.
Despite this, the axioms do not single out a unique quantum fidelity.

The remainder of this section introduces several well-known distances, fidelities, and one divergence, representing potential cost functions.

\paragraph{Hilbert-Schmidt Distance.}
The Hilbert-Schmidt inner product $\langle\rho,\sigma\rangle_\mathrm{HS} = \mytr(\rho^\dagger \sigma)$ induces a norm on $\denop(\hs)$.
This can be used to define the \textit{Hilbert-Schmidt distance},
\begin{align} \label{eq:HS_distance}
    D_\mathrm{HS}(\rho,\sigma) = \sqrt{\mytr\left( (\rho-\sigma)^2 \right)} \;,
\end{align}
where we used that $(\rho-\sigma)^\dagger = \rho-\sigma$ for $\rho,\sigma\in\denop(\hs)$.
It has a clear operational meaning as an information distance between two quantum states \cite{lee_operationally_2003}.
One advantage of this cost function choice is that it is readily measurable on a quantum computer using the SWAP test \cite{barenco_stabilization_1997, buhrman_quantum_2001}.
However, it violates the data-processing inequality \cite{ozawa_entanglement_2000}.

\paragraph{Trace Distance.}
The trace distance is given by
\begin{align} \label{eq:trace_distance}
    D_{\mytr}(\rho,\sigma) = \frac{1}{2} \mytr(|\rho - \sigma|) \;,
\end{align}
where $|A| = \sqrt{A^\dagger A}$.
It can be interpreted as follows: Given two quantum states $\rho$ and $\sigma$, each with probability $1/2$, the trace distance quantifies the lowest error probability for distinguishing them upon performing any POVM \cite{gilchrist_distance_2005}.
Furthermore, the trace distance satisfies the data-processing inequality \cite{ruskai_beyond_1994}.

\paragraph{Generalized $p$-Fidelities.}

The \textit{$p$-fidelity} \cite{liang_quantum_2019} is a general approach that covers multiple interesting similarity and distance measures.
It is defined as
\begin{align} \label{eq:p-fidelity}
    F_p(\rho,\sigma) = \frac{\Vert\sqrt{\sigma}\sqrt{\rho}\Vert_p^2}{\mathrm{max}\left(\Vert\sigma\Vert_p^2, \Vert\rho\Vert_p^2 \right)} \;,
\end{align}
where the $p$-norm is $\Vert A\Vert_p := \mathrm{tr}((A^\dagger A)^{p/2})^{1/p}$.
This satisfies all fidelity axioms for $p\geq1$.

We consider two special cases.
For $p=1$, we obtain the \textit{Uhlmann-Jozsa fidelity} \cite{jozsa_fidelity_1994, uhlmann_transition_1976}
\begin{align} \label{eq:FJ}
    F_1(\rho,\sigma)=\mytr\left(\sqrt{\sqrt{\sigma}\rho \sqrt{\sigma}}\right)^2 \;,
\end{align}
which satisfies the data-processing inequality \cite{barnum_noncommuting_1996}.
Furthermore, Uhlmann's theorem \cite{uhlmann_transition_1976} allows to connect $F_1$ to the Bures metric, a natural Riemannian metric on the space of mixed quantum states \cite{bengtsson_geometry_2006}.
The \textit{Bures distance} is given by
\begin{align} \label{eq:Bures_distance}
    D_1(\rho,\sigma) = \sqrt{2 \left(1-\sqrt{F_1(\rho,\sigma)}\right)} \;.
\end{align}
Despite having a solid theoretic foundation, $F_1$ and $D_1$ are challenging to measure experimentally.

The case $p=2$ leads to the \textit{Hilbert-Schmidt fidelity}, given by
\begin{align} \label{eq:F2}
    F_2(\rho,\sigma) = \frac{\mytr(\rho \, \sigma)}{\mathrm{max}\left\{\mytr(\rho^2), \mytr(\sigma^2) \right\}} \;.
\end{align}
Contrary to the Hilbert-Schmidt inner product, it satisfies the fidelity axioms.
The main advantage of $F_2$ is that it is easily calculable and experimentally measurable, as demonstrated in Ref.~\cite{elben_cross-platform_2020}.
However, one downside is that it violates the data-processing inequality \cite{liang_quantum_2019}.
Lastly, as shown in Ref.~\cite{liang_quantum_2019}, one can define a distance based on \eqref{eq:F2} by
\begin{align} \label{eq:D_2_distance}
    D_2(\rho,\sigma) = \sqrt{2(1 - F_2(\rho,\sigma))} \;.
\end{align}
To avoid confusion with the Hilbert-Schmidt distance \eqref{eq:HS_distance}, we refer to \eqref{eq:D_2_distance} as the $D_2$ distance.

\paragraph{Quantities from Hypothesis Testing.}

Quantum hypothesis testing is a fundamental quantum processing task where an observer receives a quantum system known to be in one of two possible states, and the goal is to correctly guess which state it is after performing a POVM measurement \cite{khatri_principles_2024}.

This setting gives rise to two fundamental asymptotic quantities.
The first one is the \textit{Quantum Chernoff Bound} \cite{audenaert_discriminating_2007, nussbaum_chernoff_2009}
\begin{align} \label{eq:QCB}
    F_\mathrm{QCB}(\rho,\sigma) = \min_{0\leq s\leq 1} \mathrm{tr}(\rho^s \sigma^{1-s}) \;.
\end{align}
Its interpretation is the following: The quantity $- \log_2 (F_\mathrm{QCB} (\rho, \sigma))$ is the optimal asymptotic error exponent for symmetric hypothesis testing, i.e., quantum state discrimination.
Additionally, it satisfies the data-processing inequality \cite{audenaert_discriminating_2007}.

The second is the \textit{quantum relative entropy}.
It is defined as
\begin{align} \label{eq:QRE}
    D_\mathrm{QRE}(\rho ||\sigma) = \mytr(\rho \log(\rho) - \rho \log(\sigma)) \;,
\end{align}
where $\log$ is the matrix logarithm.
It is a divergence, not a distance measure, as it is not symmetric under exchanging $\rho$ and $\sigma$.
Nonetheless, it can be used for state discrimination because it is non-negative, and zero if and only if $\rho=\sigma$ due to Klein's inequality \cite{klein_zur_1931}.
It also satisfies the data-processing inequality \cite{lindblad_completely_1975}.
The quantum relative entropy gains operational meaning from the quantum Stein's lemma as the optimal rate in asymmetric quantum hypothesis testing \cite{khatri_principles_2024}.

\section{Numerical Trainability Results} \label{sec:numerical_results}

To demonstrate the trainability of the extended DQNN architecture and to quantify the effect of different cost functions on the learning rate of quantum neural networks, we conduct numerical simulations of a minimal network consisting of three qubits (see Fig.~\ref{fig:minimal_universal_dqnn}).
The network implements the quantum channel $\Enet$, and its isometry parameters are initialized randomly but close to zero.
This corresponds to canonically embedding the input state in the larger Hilbert space of the whole network, with an additional small numerical perturbation.
We found that without this minor disturbance of the initial parameters, the convergence to a cost function optimum is slower.
A similar initialization strategy mitigates barren plateaus in variational quantum circuits \cite{grant_initialization_2019}.
The training objective is to learn a target quantum channel $\Etar$.
Due to the network's extended structure, the DQNN we consider is quantum channel universal for qubit-qubit channels, i.e., it can represent any such channel exactly.
This avoids the problem of $\Etar$ being impossible to learn.
The optimization is done with Choi and random state training separately (discussed in detail in Sec.~\ref{sec:optimization_of_DQNNs}).
In both cases, we use the ADAM algorithm for gradient optimization \cite{kingma_adam_2017} for 1000 training iterations.

Objective assessment of the cost functions' performance requires a suitable and independent distinguishability measure for $\Enet$ and $\Etar$.
A useful quantity is the \textit{diamond distance} $\Vert\Enet - \Etar\Vert_\diamond$ \cite{rosgen_hardness_2005}.
It is induced by the \textit{diamond norm} \cite{kitaev_quantum_1997}
\begin{align}
    \Vert\E\Vert_\diamond = \max_{\rho \in \denop(\hs\otimes\hs)} \Vert(\id_d \otimes \E)(\rho)  \Vert_1 \;,
\end{align}
where $\E$ is a CPTP map acting on $\hs$, $d=\dim(\hs)$, and $\Vert A\Vert_1 = \mytr(|A|)$ denotes the trace norm.
It can be interpreted as the best-case distinguishability of the output of the two channels when applied to part of a quantum state.
Further note that $D_{\mytr}(J(\Etar), J(\Enet)) \leq \frac{1}{2} \Vert\Enet - \Etar\Vert_\diamond$.
We employ a numerical implementation of the diamond distance using a Monte Carlo algorithm described in Ref.~\cite{benenti_computing_2010}.

In Sec.~\ref{sec:numerical_results_random}, we optimize the DQNN using randomly sampled target quantum channels, while in Sec.~\ref{sec:numerical_results_werner} we consider the highly symmetric Werner channel as the target objective.

\subsection{Learning Random Channels} \label{sec:numerical_results_random}

To determine the performance of the different cost functions, we begin the numerical analysis by training the DQNN using 100 random qubit-qubit target channels $\Etar$.
The channel sampling is implemented using \cite{gawron_quantuminformationjljulia_2018, bruzda_random_2009}.

\begin{figure*}[!ht]
    \begin{center}
    \begin{subfigure}[h]{0.48\linewidth}
    \includegraphics[width=\linewidth]{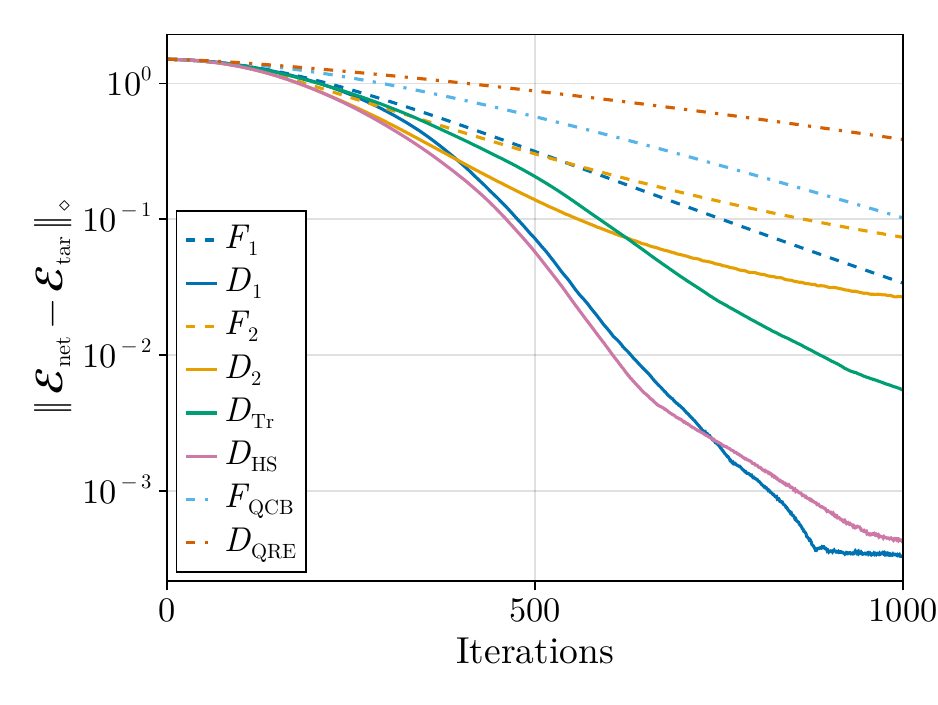}
    \caption{Mean diamond distance for Choi training.}
     \label{fig:minimal_DQNN_diamond_Choi}
    \end{subfigure}%
    \quad
    \begin{subfigure}[h]{0.48\linewidth}
    \includegraphics[width=\linewidth]{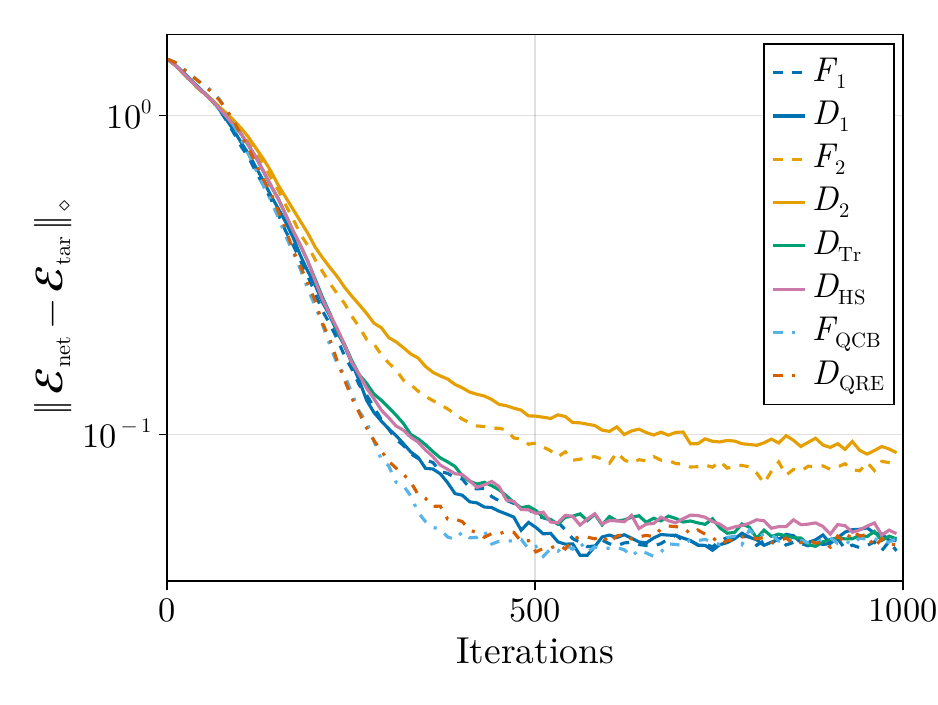}
    \caption{Mean diamond distance for random state training.}
    \label{fig:minimal_DQNN_diamond_QML}
    \end{subfigure}%
    \caption{
    The plots show the mean diamond distance $\Vert \Enet - \Etar \Vert_\diamond$ for 1000 training iterations, averaged over 100 random target channels $\Etar$.
    (a): Using Choi training, the best-performing cost functions are $D_1$ followed by $D_\mathrm{HS}$, reaching a mean diamond distance of less than $10^{-3}$.
    (b): Training the network with randomly sampled input states leads to faster convergence to a cost optimum.
    However, this optimum is worse, reaching only a mean diamond distance of around $5\times10^{-2}$ for all cost functions except $F_2$ and $D_2$.
    }
    \label{fig:minimal_DQNN_diamond}
    \end{center}
\end{figure*}

Our first benchmark comes from Choi training.
Fig.~\ref{fig:minimal_DQNN_diamond_Choi} shows the average diamond distance $\Vert \Enet - \Etar \Vert_\diamond$ for 1000 optimization iterations.
After the training, the distances $D_1$, $D_2$, $D_{\mytr}$, and $D_\mathrm{HS}$ perform better than any fidelity.
However, the learning rate of $F_1$ suggests that it may surpass $D_2$ with additional training rounds.
Nonetheless, $D_1$ and $D_{\mathrm{HS}}$ achieve the best result with a mean diamond distance of $3.43\times10^{-4}$ and $4.55\times10^{-4}$, respectively.
Interestingly, $F_\mathrm{QCB}$ and $D_\mathrm{QRE}$, both related to asymptotic hypothesis testing, lead to the least optimized networks after the training.
In these cases, the final mean diamond distance is 0.102 and 0.386, respectively.

The second benchmark is obtained using random state training.
The input training states are sampled using the Hilbert-Schmidt distribution on the set of quantum states (implemented using \cite{gawron_quantuminformationjljulia_2018, zyczkowski_induced_2001}).
For the training, we use eight batches containing four states each.
Once a cost optimum is reached for a batch, 32 new training states are generated.
Fig.~\ref{fig:minimal_DQNN_diamond_QML} shows the mean diamond distance between $\Etar$ and $\Enet$.
The convergence to the cost function optimum is faster but does not reach the same values as the Choi training.
Specifically, it converges to about $5\times10^{-2}$ for almost all examined cost functions, the exceptions being $F_2$ and $D_2$, which perform significantly worse than the others.

\subsection{Learning the Werner Channel} \label{sec:numerical_results_werner} 

Lastly, we investigate the trainability of the Werner channel, given by
\begin{align} \label{eq:Werner_channel}
    \EWalpha(\rho) = \frac{1}{\alpha + d}\left( \mytr(\rho) \, \id_d + \alpha \, \rho^T \right) \;,
\end{align}
where $\alpha\in[-1,1]$, $\rho\in\denop(\hs)$, $d=\dim(\hs)$, and $\rho^T$ denotes the transpose of $\rho$.
The name stems from the fact that the Choi state $J(\EWalpha)$ is the Werner state \cite{werner_quantum_1989}, an exceptionally symmetric bipartite quantum state with a deep connection to the foundations of quantum theory.
The Werner channel inherits many interesting features from its Choi state.
For example, it has full Kraus rank for $\alpha\in(-1,1)$, is unital, mixed-unitary for $d=2$ \cite{watrous_theory_2018}, and the output state is generally highly mixed \cite{lancien_approximating_2024}.
Furthermore, the Werner channel is entanglement breaking for $\alpha\in[-\frac{1}{d},1]$ as the Werner state is separable for this parameter region.
The case $\alpha=0$ corresponds to the completely depolarizing channel, outputting the maximally mixed state $\frac{1}{d} \id_d$ for any input state.

\begin{figure}[ht]
    \begin{center}
    \includegraphics[width=1\linewidth, keepaspectratio]{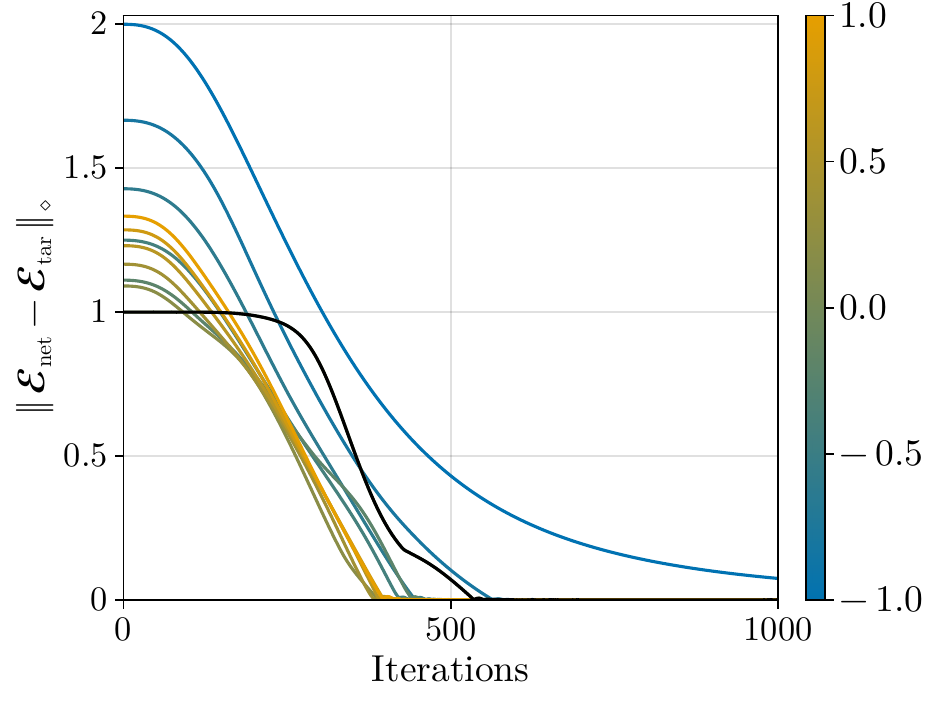}%
    \end{center}
    \vspace{-5mm}
    \caption[]{
    \label{fig:learning_Werner_channel}
    Learning the Werner channel $\EWalpha$ using Choi training with the Hilbert-Schmidt distance \eqref{eq:HS_distance} as the cost function.
    The colors indicate the value of $\alpha\in[-1,1]$.
    The completely depolarizing channel $\mathcal{E}_{\mathrm{W},0}$ is highlighted in black.
    }
\end{figure} 

Fig.~\ref{fig:learning_Werner_channel} depicts the optimization of the minimal extended DQNN using Choi training and the Hilbert-Schmidt distance cost function \eqref{eq:trace_distance} (other cost functions show similar behavior).
We find that the convergence properties correlate with $\alpha$: The higher this value, the faster the convergence to a small diamond distance.
While the diamond distance for every $\EWalpha$ with $-0.7\leq\alpha\leq1$ (except $\alpha=0$) is at most $6.7\times 10^{-3}$ after 500 training iterations, $\mathcal{E}_{\mathrm{W},-1}$ only achieves a value of $0.075$ after 1000 rounds.
Interestingly, the network seems to have problems finding an optimum in the cost landscape for $\alpha=0$: The diamond distance decreases rapidly only after around 350 optimization steps.

\section{Discussion and Conclusion} \label{sec:conclusion}

In this contribution, we developed an extension of the conventional dissipative quantum neural network architecture so that the perceptrons realize general quantum channels and investigated the impact of the cost function on the optimization process.
In particular, we found that using isometries instead of unitaries in formulating DQNNs considerably reduces the number of parameters to optimize during training.
To leverage this, we derived a versatile one-to-one composite parametrization of isometries.
Besides the established way of using randomly sampled states for training, we presented a different training method based on the network's Choi state.
The main advantage is that the optimization does not rely on the sampling method (which requires choosing a non-unique geometry of quantum states), thus allowing more objective trainability statements.
However, the target channel must be known entirely, and the cost function is applied to states with a larger Hilbert space dimension.
This Choi approach distinguishes the quantum from the classical version of feed-forward neural networks, for which random inputs are required.
We then defined a DQNN as \textit{quantum channel universal} if it can learn arbitrary quantum channels from the input to the output state.
Based on this, we argued for extending the conventional architecture by adding ancilla neurons to increase its expressivity.
This way, the individual building blocks of (large) networks are quantum channel universal at the prize of increasing their size.\\
\indent
We simulated a minimal extended network consisting of three qubits and one perceptron to evaluate the influence of different cost functions on gradient optimization.
The first objective was to learn random quantum channels to obtain insight into the general convergence behavior.
Using Choi training, we found that the Hilbert-Schmidt and Bures distance performed best.
Due to the fact that the former is easily calculable and readily measurable on quantum hardware, we suggest this to be the preferred cost function for Choi training.
Furthermore, as its computation only involves functions that are at most quadratic in the quantum states, shadow tomography via randomized measurements \cite{aaronson_shadow_2018, huang_predicting_2020} is an alternative to full quantum state tomography \cite{mauro_dariano_quantum_2003, guta_fast_2020}.
For random state training, almost all cost functions performed equally well.
Nonetheless, for the Hilbert-Schmidt and Bures distance, the final distinguishability between the network and target channel was about two orders of magnitude greater than for Choi training.
However, the convergence to an optimum is faster than for Choi training.
This can be interpreted as the DQNN showing signs of barren plateaus for Choi training (i.e., gradients that vanish exponentially with the Hilbert space dimension) due to the cost function acting on a larger Hilbert space \cite{cerezo_cost_2021, sharma_trainability_2022}.
This well-known trainability issue is not exclusive to gradient-based optimization, which we used in this work, but also appears in gradient-free schemes \cite{arrasmith_effect_2021}.
The presented results suggest that this phenomenon does not affect all cost functions equally (compare, e.g., the differences between Choi and random state training for the Hilbert-Schmidt distance and the quantum relative entropy, respectively).
This raises the question of what properties a cost function must have to be less prone to barren plateaus.
Our findings indicate that satisfying the data-processing inequality is not the decisive factor.\\
\indent
Lastly, we studied the trainability of the Werner channel.
The results indicate a correlation between the learning rate and the Werner channel's parameter.
This suggests a connection between the trainability and the target channel's properties.
For example, the optimization takes less iterations if the channel is entanglement breaking.
The exception is the completely depolarizing channel, for which the network has initial difficulties finding an optimum in the cost function landscape.\\
\indent
In conclusion, our results shed new light on two crucial aspects of quantum neural network design: architecture and cost function.
We believe that the isometry formulation of extended DQNNs will aid in the theoretical development of this growing field.
Furthermore, having found a suitable and readily measurable cost function will influence the experimental realization of quantum machine learning models.

\section*{Acknowledgments}

T.C.S. wants to thank Felix Hitzelhammer for valuable discussions and comments.
B.C.H. and C.P. acknowledge gratefully that this research was funded in whole, or in part, by the Austrian Science Fund (FWF) project P36102-N (Grant DOI:  10.55776/P36102).
For the purpose of open access, the author has applied a CC BY public copyright license to any Author Accepted Manuscript version arising from this submission. 
The funder played no role in study design, data collection, analysis and interpretation of data, or the writing of this manuscript.






\bibliographystyle{quantum}
\bibliography{references}

\onecolumn
\appendix

\section{The Composite Parametrization of Isometries} \label{app:comp_param_for_isometries}

In this section, we derive the composite parametrization for isometries.
As the name suggests, it is obtained from the composite parametrization of the unitary group $\mathcal{U}(d)$ introduced in Ref.~\cite{spengler_composite_2010, spengler_composite_2012}.

Let $\hs_1$ and $\hs_2$ be Hilbert spaces with dimensions $d_1$ and $d_2$, respectively.
Let $\linop(\hs)$ be the set of linear operators mapping $\hs$ into itself, and $\mathrm{Iso}(\hs_1,\hs_2)$ the set of isometries from $\hs_1$ to $\hs_2$ with $d_1\leq d_2$.
Furthermore, let $\{|i\rangle_\ell\}_{i=0}^{d_\ell -1}$ be the computational basis of $\hs_\ell$.

Any isometry $V\in\mathrm{Iso}(\hs_1 , \hs_2)$ with $d_1\leq d_2$ can be written as
\begin{align} \label{eq:cptp_stinespring_comp_param_iso_0}
    V = U\; \id_{d_{2}\times d_1} \;,
\end{align}
where $U\in\U(d_{2})$, and $\id_{d_{2}\times d_1} \in \mathrm{Iso}(\hs_1, \hs_2)$ denotes the $d_{2}\times d_1$ matrix consisting of the first $d_1$ columns of $\id_{d_{2}\times d_{2}}$.
It can be written as
\begin{align}
    \id_{d_{2}\times d_1} = \sum_{i=0}^{d_1-1} |i\rangle_{2} \langle i|_1 \;.
\end{align}
Note that not all basis vectors $|i\rangle_{2}$ of $\hs_{2}$ need to appear in this sum.
By \eqref{eq:cptp_stinespring_comp_param_iso_0}, the action $V A V^\dagger$ of $V$ on $A\in\linop(\hs_1)$ can be interpreted as first canonically embedding $A$ into $\linop(\hs_2)$ and subsequently applying the unitary $U$.

We can use the composite parametrization of unitary matrices \cite{spengler_composite_2010, spengler_composite_2012} to write $U$ as
\begin{align} \label{eq:unitary_comp_param}
    U = \left[\prod_{m=0}^{d_{2}-2} \left(\prod_{n=m+1}^{d_{2}-1} \Lambda_{m,n}  \right) \right]\,\left[\prod_{l=0}^{d_{2}-1} e^{i P_l \lambda_{ll}} \right] \;,
\end{align}
with
\begin{align}
    P_n &:= |n\rangle_{2} \langle n|_{2} \;, \label{eq:P_n}\\
    Y_{m,n} &:= -i |m\rangle_{2}\langle n|_{2} + i |n\rangle_{2}\langle m|_{2} \;\; , \; 0\leq m<n\leq d_{2}-1 \;, \label{eq:Y_mn}\\
    \Lambda_{m,n} &:= e^{i\, P_n \lambda_{n,m}} e^{i\, Y_{m,n} \lambda_{m,n}} \;,
\end{align}
and $\lambda_{m,n} \in [0,2\pi]$ for $m\geq n$ and $\lambda_{m,n} \in [0,\pi/2]$ for $m<n$.

Using \eqref{eq:cptp_stinespring_comp_param_iso_0} and \eqref{eq:unitary_comp_param} we compute
\begin{align}
    \prod_{l=0}^{d_{2}-1} e^{i P_l \lambda_{ll}} \; \id_{d_{2}\times d_1} &= \left(\sum_{l=0}^{d_{2}-1} e^{i\, \lambda_{l,l}} |l\rangle_{2} \langle l|_{2}\right) \left( \sum_{k=0}^{d_1-1} |k\rangle_{2} \langle k|_1 \right) \\
    &= \sum_{k=0}^{d_1-1} e^{i\, \lambda_{k,k}} |k\rangle_{2} \langle k|_1 \;, \label{eq:cptp_stinespring_comp_param_1}
\end{align}
and observe that only the first $d_1$ phases $\lambda_{k,k}$ are relevant for the isometry.
This $d_{2}\times d_1$-dimensional matrix is explicitly given by
\begin{align} \label{eq:cptp_stinespring_comp_param_1_matrixform}
    \prod_{l=0}^{d_{2}-1} e^{i P_l \lambda_{ll}} \; \id_{d_{2}\times d_1} = 
        \left( {\begin{array}{ccc}
            e^{i\, \lambda_{0,0}} & 0  & 0 \\
            0 & \ddots & 0 \\
            0 & \cdots & e^{i\, \lambda_{d_1-1, d_1-1}} \\
            \vdots & & \vdots \\
            0 & \cdots & 0
    \end{array} } \right) \;,
\end{align}
Next, we calculate
\begin{align}
    \Lambda_{m,n} =& \id_{d_{2}\times d_{2}} + (c_{m,n} - 1 ) P_m + (e_{n,m} c_{m,n} -1) P_n - e_{n,m} s_{m,n} |n\rangle_{2} \langle m|_{2} + s_{m,n} |m\rangle_{2} \langle n|_{2} \;,
\end{align}
where we abbreviated $\sin{\lambda_{m,n}} = s_{m,n}$, $\cos{\lambda_{m,n}} = c_{m,n}$, and $e^{i\lambda_{n,m}} = e_{n,m}$.
In matrix notation, this amounts to
\begin{align}
  \Lambda_{m,n} =
  \left( {\begin{array}{ccccc}
    \id_{{m} \times {m}} & 0 & 0 & 0 & 0\\
    0 & c_{m,n} & 0 & s_{m,n} & 0\\
    0 & 0 & \id_{(n-m-1)\times (n-m-1)} & 0 & 0 \\
    0 & - e_{n,m}\, s_{m,n} & 0 & e_{n,m}\, c_{m,n} & 0 \\
    0 & 0 & 0 & 0 & \id_{(d_{2}-n-1)\times (d_{2}-n-1)}
  \end{array} } \right) \;.
\end{align}
By inspection, we see that $\Lambda_{m,n}$ acts trivially from the left on a matrix of the form \eqref{eq:cptp_stinespring_comp_param_1_matrixform} if $m\geq d_1$.
Thus, we can write the isometry \eqref{eq:cptp_stinespring_comp_param_iso_0} as
\begin{align}
    V &= \left[\prod_{m=0}^{d_{2}-2} \left(\prod_{n=m+1}^{d_{2}-1} \Lambda_{m,n}  \right) \right]\,\left[\prod_{l=0}^{d_{2}-1} e^{i P_l \lambda_{ll}} \right] \; \id_{d_{2}\times d_1} \\
    &= \left[\prod_{m=0}^{d_1-1} \left(\prod_{n=m+1}^{d_{2}-1} \Lambda_{m,n}  \right) \right]\,\left[\prod_{l=0}^{d_1-1} e^{i P_l \lambda_{ll}} \right] \; \id_{d_{2}\times d_1} \;. \label{eq:cptp_stinespring_comp_param_final}
\end{align}

Regarding the parameter count in \eqref{eq:cptp_stinespring_comp_param_final}, we note that the first term on the right-hand side gives $\sum_{m=0}^{d_1-1} 2 (d_{2}-m-1) = 2d_1 d_{2} - d_1^2 - d_1$ because each $\Lambda_{m,n}$ introduces two degrees of freedom.
The second term on the right-hand side gives an additional $d_1$ free parameters.
Thus, we find that the isometry $V\in\mathrm{Iso}(\hs_1 , \hs_2)$ in \eqref{eq:cptp_stinespring_comp_param_final} has $2d_1 d_{2} - d_1^2$ free real parameters.
A simple argument shows that this is indeed the number of free real parameters of a general isometry in $\mathrm{Iso}(\hs_1,\hs_2)$.
Consequently, we cannot eliminate more parameters from \eqref{eq:cptp_stinespring_comp_param_final}.

The parameters of the isometry \eqref{eq:cptp_stinespring_comp_param_final} can be conveniently collected in the matrix
\begin{align} \label{eq:cptp_stinespring_comp_param_param_matrix}
  (\lambda_{m,n})_{m,n} =
  \left( {\begin{array}{cccccc}
    \lambda_{0,0} & \cdots & \lambda_{0,d_1-1} & \cdots &\cdots & \lambda_{0,d_{2}-1} \\
    \vdots & \ddots & \vdots & & & \vdots \\
    \lambda_{d_1-1,0} & \cdots & \lambda_{d_1-1, d_1-1} & \cdots & \cdots & \lambda_{d_1-1, d_{2}-1} \\
    \vdots & & \vdots & 0 & \cdots & 0 \\
    \vdots & & \vdots & \vdots & & \vdots \\
    \lambda_{d_{2}-1,0} & \cdots & \lambda_{d_{2}-1,d_1-1} & 0 & \cdots & 0
  \end{array} } \right) \;,
\end{align}

The diagonal entries $\lambda_{n,n}$ correspond to global phases in the respective subspaces.
The entries $\lambda_{m,n}$ in the upper triangular part represent rotations in the subspaces spanned by $|n\rangle_2$ and $|m\rangle_2$, and $\lambda_{n,m}$ in the lower triangular part correspond to relative phases in these subspaces.
Except for the diagonal entries $\lambda_{k,k}$, these are the same parameters needed to parameterize a general mixed state $\rho_{2} \in \denop(\hs_{2})$ of rank $d_1\leq d_{2}$ (in addition to $d_1-1$ required mixing probabilities; cf. \cite{spengler_composite_2010}).

\subsection{Composite Parametrization for the Stinespring Isometry of a Quantum Channel}

A general CPTP map $\mathcal{E}:\denop(\hs_1)\rightarrow\denop(\hs_2)$ can be written in its Stinespring representation as
\begin{align} \label{eq:cptp_stinespring}
    \mathcal{E}(\rho) = \mytr_{1,A}(V \rho V^\dagger) \;,
\end{align}
where $V\in\mathrm{Iso}(\hs_1,\hs_2\otimes\hs_A\otimes\hs_1)$, and the Hilbert space $\hs_A$ with $\dim(\hs_A) = d_A$ corresponds to an ancilla system.
We can choose $d_A = d_2$ because $d_1 d_2$ is the maximal Kraus rank of $\mathcal{E}$ \cite{watrous_theory_2018}.
In this case, the isometry in \eqref{eq:cptp_stinespring_comp_param_final} with $d_2 d_A d_1 = d_1 d_2^2$ has $d_1^2 (2 d_2^2 - 1)$ degrees of freedom.
However, due to the unitary freedom on the space $\hs_A \otimes \hs_1$ (which has no physical relevance), a general CPTP map can be reduced to $d_1^2 (d_2^2 - 1)$ degrees of freedom.
Unfortunately, we cannot straightforwardly get rid of the $d_1^2 d_2^2$ redundant parameters in \eqref{eq:cptp_stinespring_comp_param_param_matrix} because these degrees of freedom do not coincide one-to-one with the parameters $\lambda_{i,j}$.

Nonetheless, using \eqref{eq:cptp_stinespring_comp_param_final}, we can write a Stinespring isometry for $\mathcal{E}$ as
\begin{align} \label{eq:cptp_stinespring_channel_isometry}
    V = U_C \; \id_{d_1 d_2^2\times d_1} = U_C \; \left(|0\rangle_2 \otimes |0\rangle_A \otimes \id_1 \right) \;,
\end{align}
with
\begin{align} \label{eq:cptp_unitary}
    U_C = \left[\prod_{m=0}^{d_1-1} \left(\prod_{n=m+1}^{d_1 d_2^2-1} \Lambda_{m,n}  \right) \right]\,\left[\prod_{l=0}^{d_1-1} e^{i P_l \lambda_{ll}} \right] \;.
\end{align}
We thus obtain for any $\rho_1 \in \denop(\hs_1)$:
\begin{align}
    \E (\rho_1) &= \mytr_{1,A}[V \, \rho_1 \, V^\dagger] \\
    &= \sum_{k=0}^{d_2-1} \sum_{l=0}^{d_1-1} \langle k|_A \otimes \langle l|_1 \, V \; \rho_1 \; V^\dagger\, |k\rangle_A \otimes |l\rangle_1 \\
    &= \sum_{k=0}^{d_2-1} \sum_{l=0}^{d_1-1} \langle k|_A \otimes \langle l|_1 \,U_C \,|0\rangle_2 \otimes |0\rangle_A \; \rho_1 \; \langle 0|_2 \otimes \langle 0|_A\, U_C^\dagger\, |k\rangle_A \otimes |l\rangle_1 \\
    &= \sum_{k=0}^{d_2-1} \sum_{l=0}^{d_1-1} G_{k,l} \,\rho_1\, G_{k,l}^\dagger \;,
\end{align}
with the Kraus operators $G_{k,l} =\langle k|_A \otimes \langle l|_1 V = \langle k|_A \otimes \langle l|_1 U_C |0\rangle_2 \otimes |0\rangle_A$.

Note that this representation of a quantum channel $\mathcal{E}$ requires us to tensor the systems $\hs_A$ and $\hs_2$ from the left onto $\hs_1$ to obtain $|0\rangle_2\otimes|0\rangle_A\otimes\id_1$ in \eqref{eq:cptp_stinespring_channel_isometry}.
In this case, the parameters in the composite parametrized isometry are reduced to the correct number $d_1^2(2 d_2^2-1)$.
If we tensor the systems $\hs_A$ and $\hs_2$ from the right onto $\hs_1$, we would get $\id_1\otimes|0\rangle_A\otimes|0\rangle_2$ in \eqref{eq:cptp_stinespring_channel_isometry}.
Consequently, the $d_1^2 d_2^4$ parameters of the full unitary $U_C$ get reduced only by less than $d_1^2 (d_2^2 -1)^2$, and \eqref{eq:cptp_unitary} is no longer valid.
The reason is that the reduction of parameters shown above cannot be carried out.
This also becomes apparent when numerically optimizing DQNNs using the composite parametrization because the gradient for redundant degrees of freedom in the unitary formalism vanishes.
If tensored in the ``wrong'' order, only some diagonal elements of the parameter matrix \eqref{eq:cptp_stinespring_comp_param_param_matrix} are irrelevant for the quantum channel and have vanishing derivative (cf.~\cite{hiesmayr_quantum_2024}).
If done correctly, only the parameters $\lambda_{i,j}$ in \eqref{eq:cptp_stinespring_comp_param_param_matrix} are relevant for the optimization, i.e., have non-vanishing derivative in general.
Consequently, the derivative of the cost function only needs to be calculated for those, leading to better computational performance.

\section{Derivatives of different cost functions} \label{app:derivatives_cost_funct}

This section presents the derivatives of the different cost functions required for optimizing a DQNN by gradient descent/ascent.
In Sec.~\ref{sec:numerical_results}, the resulting gradient matrix is used to update the isometry parameters, e.g., with the ADAM optimizer \cite{kingma_adam_2017}.

For example, consider an extended DQNN consisting of 5 qudits and two perceptrons $U_1$ and $U_2$ in the unitary formulation.
It comprises one input, one hidden, one output, and two ancilla layers.
Using the notation of Sec.~\ref{sec:extended_DQNNs}, we have $U_1 \equiv U_1^{(1,2,3)}$ and $U_2 \equiv U_1^{(3,4,5)}$.
Let us denote the set of variational parameters in $U_m$ as $\mathcal{S}_m = \{\lambda_{x,y}^{(m)}\}_{x,y}$.
Due to the equivalence of the isometry and the unitary picture, this is already a reduced set of parameters (cf. \eqref{eq:cptp_stinespring_comp_param_param_matrix}).
The network channel is given by
\begin{align}
    \rhoout = \Enet(\rhoin) = \mytr_{1,2,3,4}\left\{ \, U \, (|\mathbf{0}\rangle\langle\mathbf{0}|\otimes\rhoin) \, U^\dagger \, \right\} \;,
\end{align}
where $U = U_2 \, U_1$, and $|\mathbf{0}\rangle\langle\mathbf{0}| \equiv |\mathbf{0}\rangle\langle\mathbf{0}|_{5,4,3,2}$ is the initial state of the ancilla, hidden, and output layers.

Gradient optimization utilizes either the numerical or the analytical derivative of the total cost function \eqref{eq:cost_total}.
In the former case, we approximate the cost function gradient by
\begin{align}
     \frac{\partial C_\mathrm{tot}}{\partial \lambda_{x,y}^{(m)}} 
    &= \frac{1}{N_t} \sum_{i=1}^{N_t} \frac{\partial}{\partial \lambda_{x,y}^{(m)}} C(\rho_\mathrm{tar}^{(i)}, \, \rho_\mathrm{out}^{(i)}) \label{eq:d_C_tot}\\
    &\approx \frac{1}{N_t} \sum_{i=1}^{N_t} \frac{C(\rhotar^{(i)},\,\rhoout^{(i)}(\lambda_{x,y}^{(m)}+\varepsilon)) - C(\rhotar^{(i)},\,\rhoout^{(i)}(\lambda_{x,y}^{(m)}-\varepsilon))}{2\varepsilon}
\end{align}
for each $m\in\{1,2\}$ and relevant $x,y$ (cf.~\eqref{eq:cptp_stinespring_comp_param_param_matrix}).
This method requires choosing a suitable small $\varepsilon>0$ and is generally only an approximation of the true gradient.
Thus, we resort to it only when we cannot compute the cost function's analytical derivative.
This is the case for the quantum Chernoff bound \eqref{eq:QCB} and the quantum relative entropy \eqref{eq:QRE}.

For the analytic approach, we compute \eqref{eq:d_C_tot} analytically for each $m\in\{1,2\}$ and relevant $x,y$.
To unclutter the notation in the following, we focus on one input-target pair of the training set (i.e., one term in the sum \eqref{eq:d_C_tot}) and drop the superscript $(i)$.
When evaluating $\partial C(\rhotar,\rhoout) / \partial \lambda_{x,y}^{(m)}$, we necessarily encounter $\partial \rhoout / \partial \lambda_{x,y}^{(m)}$ due to the chain rule.
Thus, before moving on, we first take care of this.
As shown in \cite{hiesmayr_quantum_2024}, we can calculate for our example DQNN
\begin{align}
    \frac{\partial \rhoout}{\partial \lambda_{x,y}^{(1)}} &= \mytr_{1,2,3,4}\left\{\, U_2\, U_1\, i\, [\, \Tilde{Y}_{x,y}^{(1)},\, \Tilde{\rho} \,]\, U_1^\dagger\, U_2^\dagger\,  \right\} \;, \label{eq:drho_1}\\
    \frac{\partial \rhoout}{\partial \lambda_{x,y}^{(2)}} &= \mytr_{1,2,3,4}\left\{\, U_2\, i\, [\, \Tilde{Y}_{x,y}^{(2)},\, U_1\, \Tilde{\rho} \, U_1^\dagger \,]\, U_2^\dagger\,  \right\} \label{eq:drho_2} \;,
\end{align}
where $[\cdot,\cdot]$ denotes the commutator, $\Tilde{\rho} = |\mathbf{0}\rangle\langle\mathbf{0}|\otimes\rhoin$, and
\begin{align}
    \Tilde{Y}_{x,y}^{(m)} =
    \begin{cases}
        U_m^\dagger \, Y_{x,y}^{(m)} \, U_m \; &, x<y \\
        P_x^{(m)}  &, x=y \\
        U_m^\dagger \, P_x^{(m)} \, U_m  &, x>y
    \end{cases}
\end{align}
with $P_x^{(m)}$ and $Y_{x,y}^{(m)}$ as in \eqref{eq:P_n} and \eqref{eq:Y_mn}, respectively.
Generalizing the calculation for \eqref{eq:drho_1} and \eqref{eq:drho_2} to larger networks with more (unitary) perceptrons $U_m$ is straightforward.

The remainder of this section deals with evaluating $\partial C(\rhotar,\rhoout) / \partial \lambda_{x,y}^{(m)}$ for the cost functions presented in Sec.~\ref{sec:cost_funct}.
It is important to keep in mind that only $\rhoout$ depends on the variational parameters $\lambda_{x,y}^{(m)}$.
Hence, $\partial\rhoout / \partial \lambda_{x,y}^{(m)} \neq 0$ in general, but $\partial \rhotar / \partial \lambda_{x,y}^{(m)} \equiv0$.
Throughout, we furthermore assume $\rhotar \neq\rhoout$ as this can become problematic for the gradient of certain cost functions.
This condition can be implemented in the training algorithm: Before computing the gradient for any optimization iteration, check if $\rhoout = \rhotar$.
If yes, the network is already optimal for this training state, and we exclude this instance from the present iteration of the gradient calculation.

\subsection{Derivative of $D_\mathrm{HS}$}
A straightforward calculation using \eqref{eq:HS_distance} yields
\begin{align}
    \frac{\partial D_\mathrm{HS}(\rhotar,\rhoout)}{\partial \lambda_{x,y}^{(m)}}
    &= \frac{\partial}{\partial \lambda_{x,y}^{(m)}} \sqrt{\mytr\left( (\rhoout-\rhotar)^2 \right)} \\
    &= \frac{1}{2 D_\mathrm{HS}(\rhotar,\rhoout)} \frac{\partial}{\partial \lambda_{x,y}^{(m)}} \mytr\left( (\rhoout-\rhotar)^2 \right) \\
    &= \frac{1}{D_\mathrm{HS}(\rhotar,\rhoout)} \mytr\left((\rhoout - \rhotar) \, \frac{\partial \rhoout}{\partial \lambda_{x,y}^{(m)}} \right) \;.
\end{align}

\subsection{Derivative of $D_{\mytr}$}
The derivative of the trace distance cost function \eqref{eq:trace_distance} is
\begin{align}
    \frac{\partial D_{\mytr}(\rhotar,\rhoout)}{\partial \lambda_{x,y}^{(m)}}
    &= \frac{1}{2} \frac{\partial}{\partial \lambda_{x,y}^{(m)}} \mytr\left( \sqrt{(\rhoout-\rhotar)^2} \right) \\
    &= \frac{1}{4} \mytr\left( \left( \sqrt{(\rhoout-\rhotar)^2} \right)^{-1} \, \frac{\partial}{\partial \lambda_{x,y}^{(m)}} (\rhoout-\rhotar)^2 \right) \label{eq:d_D_tr_2} \\
    &= \frac{1}{2} \mytr\left( \left( \sqrt{(\rhoout-\rhotar)^2} \right)^{-1} \, (\rhoout-\rhotar) \, \frac{\partial \rhoout}{\partial \lambda_{x,y}^{(m)}} \right) \;.
\end{align}
If $A = \sqrt{(\rhoout-\rhotar)^2}$ is singular, we use the Moore-Penrose inverse \cite{penrose_generalized_1955} instead of $A^{-1}$.
In more detail, in \eqref{eq:d_D_tr_2} we used
\begin{align} \label{eq:d_sqrt_A}
    \mytr \left( \frac{\partial}{\partial\lambda} \sqrt{A} \right) = \mytr \left( \frac{1}{2} \, \frac{\partial A}{\partial\lambda} \, \left(A^{1/2}\right)^{-1} \right)
\end{align}
for $\lambda=\lambda_{x,y}^{(m)}$ and $A=(\rhoout-\rhotar)^2$.
To evaluate this, we can utilize the power series expansion for the matrix square root, given by
\begin{align} \label{eq:matrix_sqrt_power_series}
    A^{1/2} = \sum_{n=0}^\infty (-1)^n \binom{1/2}{n} (\id - A)^n \;,
\end{align}
which is convergent if the spectrum of $A$ satisfies $spec(A) \subseteq \mathcal{D}(1,1) \subset \mathbb{C}$, where $\mathcal{D}(1,1)$ denotes a disk with radius 1 and centered at 1 in $\mathbb{C}$.
If $spec(A) \subseteq (0,1]$, i.e., $A$ is non-singular, the inverse of $\eqref{eq:matrix_sqrt_power_series}$ is
\begin{align} \label{eq:matrix_sqrt_power_series_inverse}
    A^{-1/2} = \sum_{n=0}^\infty (-1)^n \binom{-1/2}{n} (\id - A)^n \;.
\end{align}
For singular $A$ we can use the Moore-Penrose pseudoinverse to define \eqref{eq:matrix_sqrt_power_series_inverse}.
We can then calculate
\begin{align}
    \mytr \left( \frac{\partial}{\partial\lambda} \sqrt{A} \right)
    &= \mytr \left( \frac{\partial}{\partial\lambda} \sum_{n=0}^\infty (-1)^n \binom{1/2}{n} (\id - A)^n \right) \\
    &= \mytr \left( \sum_{n=0}^\infty (-1)^n \binom{1/2}{n} \sum_{k=0}^{n-1} (\id - A)^k \frac{\partial (\id - A)}{\partial\lambda} (\id - A)^{n-k-1} \right) \\
    &= \mytr \left( \frac{\partial A}{\partial\lambda} \sum_{n=1}^\infty (-1)^{n+1} \binom{1/2}{n} \, n \, (\id - A)^{n-1} \right) \\
    &= \mytr \left( \frac{1}{2} \, \frac{\partial A}{\partial\lambda} \, A^{-1/2} \right) \;,
\end{align}
where in the third equality we used that the trace is cyclic, and a simple index shift in combination with the identity $\binom{1/2}{n+1}\,(n+1) = \frac{1}{2} \, \binom{-1/2}{n}$ yields the last line.

\subsection{Derivative of $F_1$ and $D_1$}

The derivative of the fidelity cost function \eqref{eq:FJ} takes the form
\begin{align}
    \frac{\partial F_1(\rhotar, \rhoout )}{\partial \lambda_{x,y}^{(m)}}
    &= \frac{\partial}{\partial \lambda_{x,y}^{(m)}} \mytr\left( \sqrt{\sqrt{\rhotar}\, \rhoout\, \sqrt{\rhotar}} \right)^2 \\
    &= 2 \sqrt{F_1(\rhotar, \rhoout )} \, \mytr\left( \frac{\partial}{\partial \lambda_{x,y}^{(m)}} \sqrt{\sqrt{\rhotar}\, \rhoout\, \sqrt{\rhotar}} \right) \\
    &= \sqrt{F_1(\rhotar, \rhoout )} \, \mytr\left( \frac{\partial \rhoout}{\partial \lambda_{x,y}^{(m)}}\, \sqrt{\rhotar} \left(\sqrt{ \sqrt{\rhotar}\, \rhoout\, \sqrt{\rhotar} }\right)^{-1} \sqrt{\rhotar} \right) \label{eq:d_F_1}\;,
\end{align}
where in the last line we used \eqref{eq:d_sqrt_A} with $A = \sqrt{\rhotar}\, \rhoout\, \sqrt{\rhotar}$.
This is valid as one can show that $spec(\sqrt{\rhotar}\, \rhoout\, \sqrt{\rhotar}) \subset [0,1]$.

For the Bures distance cost function \eqref{eq:Bures_distance}, we obtain
\begin{align}
    \frac{\partial D_1(\rhotar, \rhoout )}{\partial \lambda_{x,y}^{(m)}}
    &= \frac{\partial}{\partial \lambda_{x,y}^{(m)}} \sqrt{2 \left(1-\sqrt{F_1(\rhotar,\rhoout)}\right)} \\
    &= \frac{1}{D_1(\rhotar, \rhoout )} \frac{\partial }{\partial \lambda_{x,y}^{(m)}} \left(1-\sqrt{F_1(\rhotar,\rhoout)}\right) \\
    &= \frac{-1}{2 D_1(\rhotar, \rhoout ) \sqrt{F_1(\rhotar,\rhoout)}} \frac{\partial F_1(\rhotar, \rhoout )}{\partial \lambda_{x,y}^{(m)}} \\
    &= \frac{-1}{2 D_1(\rhotar, \rhoout )} \mytr\left( \frac{\partial \rhoout}{\partial \lambda_{x,y}^{(m)}}\, \sqrt{\rhotar} \left(\sqrt{ \sqrt{\rhotar}\, \rhoout\, \sqrt{\rhotar} }\right)^{-1} \sqrt{\rhotar} \right) \;,
\end{align}
where we used \eqref{eq:d_F_1} in the last line.

\subsection{Derivative of $F_2$ and $D_2$}

To take the derivative of the fidelity $F_2$, we write it as
\begin{align}
    F_2(\rhotar,\rhoout) = \frac{\mathcal{A}}{\mathcal{B}} \,
\end{align}
where
\begin{align}
    \mathcal{A} &= \mytr(\rhoout \rhotar) \;,\\
    \mathcal{B} &= \max\left\{\mytr(\rhoout^2), \mytr(\rhotar^2)\right\} \;.
\end{align}
Thus, we have
\begin{align} \label{eq:d_F_2_raw}
    \frac{\partial F_2(\rhotar, \rhoout )}{\partial \lambda_{x,y}^{(m)}} = \frac{\frac{\partial \mathcal{A}}{\partial \lambda_{x,y}^{(m)}} \mathcal{B} - \mathcal{A} \frac{\partial \mathcal{B}}{\partial \lambda_{x,y}^{(m)}}}{\mathcal{B}^2} \;.
\end{align}
The derivative of $\mathcal{A}$ is easily evaluated to be
\begin{align} \label{eq:d_A_for_F_2}
    \frac{\partial \mathcal{A}}{\partial \lambda_{x,y}^{(m)}} = \mytr\left( \frac{\partial \rhoout}{\partial \lambda_{x,y}^{(m)}} \, \rhotar \right) \;.
\end{align}
By writing
\begin{align}
    \max\left\{\mytr(\rhoout^2), \mytr(\rhotar^2)\right\} &= \frac{1}{2} \left( \mytr(\rhoout^2) + \mytr(\rhotar^2) + \left| \mytr(\rhoout^2) - \mytr(\rhotar^2)  \right| \right) \\
    &= \frac{1}{2} \left( \mytr(\rhoout^2 + \rhotar^2) + \sqrt{ \mytr(\rhoout^2 - \rhotar^2)^2}  \right) \;,
\end{align}
we see that $\mathcal{B}$ is not differentiable at $\rhoout=\rhotar$.
For all other cases, we can compute
\begin{align}
    \frac{\partial \mathcal{B}}{\partial \lambda_{x,y}^{(m)}} 
    &= \mytr\left( \frac{\partial \rhoout}{\partial \lambda_{x,y}^{(m)}} \,\rhoout \right) \left( 1+\frac{\mytr(\rhoout^2 - \rhotar^2)}{ \sqrt{ \mytr(\rhoout^2 - \rhotar^2)^2}} \right) \\
    &= \mytr\left( \frac{\partial \rhoout}{\partial \lambda_{x,y}^{(m)}} \,\rhoout \right) \left( 1+\mathrm{sign}\left(\mytr(\rhoout^2 - \rhotar^2)\right) \right) \label{eq:d_B_for_F_2}\;.
\end{align}
Inserting \eqref{eq:d_A_for_F_2} and \eqref{eq:d_B_for_F_2} into \eqref{eq:d_F_2_raw} finally yields after some simplifications
\begin{align} \label{eq:d_F_2}
    \frac{\partial F_2(\rhotar, \rhoout )}{\partial \lambda_{x,y}^{(m)}} = \frac{
    \mytr\left( \frac{\partial \rhoout}{\partial \lambda_{x,y}^{(m)}} \left(\rhotar - \rhoout F_2(\rhotar,\rhoout) \left( 1+\mathrm{sign}\left(\mytr(\rhoout^2 - \rhotar^2)\right) \right) \right)\right)
    }{
    \max\left\{\mytr(\rhoout^2), \mytr(\rhotar^2)\right\}
    } \;.
\end{align}

For the $D_2$ distance \eqref{eq:D_2_distance}, the derivative evaluates to
\begin{align}
    \frac{\partial D_2(\rhotar, \rhoout )}{\partial \lambda_{x,y}^{(m)}} &= \frac{-1}{D_2(\rhotar, \rhoout )} \frac{\partial F_2(\rhotar, \rhoout )}{\partial \lambda_{x,y}^{(m)}} \\
    &= \frac{
    -\mytr\left( \frac{\partial \rhoout}{\partial \lambda_{x,y}^{(m)}} \left(\rhotar - \rhoout F_2(\rhotar,\rhoout) \left( 1+\mathrm{sign}\left(\mytr(\rhoout^2 - \rhotar^2)\right) \right) \right)\right)
    }{
    D_2(\rhotar, \rhoout ) \max\left\{\mytr(\rhoout^2), \mytr(\rhotar^2)\right\}
    } \;.
\end{align}

\end{document}